\documentstyle[12pt]{article}
\begin{document}

\def\TITLE{%
Exact charge conservation scheme for Particle-in-Cell simulations
for a big class of form-factors
}

\def\TIT{%
Exact charge conservation for PIC
}

{\Large\bf
\begin{center}
\TITLE
\end{center}
}

\begin{center}

{\bf
T.Zh.Esirkepov
}

\bigskip
{Forum for Theoretical Physics INFM, Pisa, Italy}

{Moscow Institute of Physics and Technology, Institutskij per.9,
Dolgoprudnij, Moscow region, 141700 Russia}

tel. \& fax.: +7 (095) 4086772

e-mail: timur@nonlin-gw.rphys.mipt.ru
\end{center}

\vfill
\paragraph{Subject classifications:}
{\bf
65C20 Models, numerical methods;
65P20 Solution of discretized equations;
70F10 $n$-body problem;
77F05 Fluid-particle models.
}


\bigskip
~

%%%%%%%%%%%%%%%%%%%%%%%%%%%%%%%%%%%%%%%%%%%%%%%%%%%
\newpage

\begin{quotation}
\noindent
Abstract.
As an alternative to solving of Poisson equation
in Particle-in-Cell methods,
a new construction of current density
exactly satisfying continuity equation
in finite differences is developed.
This procedure called density decomposition
is proved to be the only possible
linear procedure for defining the current density
associated with the motion of a particle.
Density decomposition is valid at least for any n-dimensional form-factor
which is the product of one-dimensional form-factors.
The algorithm is demonstrated for parabolic spline form-factor.
\end{quotation}

%%%%%%%%%%%%%%%%%%%%%%%%%%%%%%%%%%%%%%%%%%%%%%%%%%%%%%%%%%%%

\section{Introduction}

In the present paper we develope a new procedure called density decomposition
for obtaining the current density automatically satisfying
the continuity equation.

In the set of Maxwell equations
along with hyperbolic equations of wave propagation
we have an equation of elliptic type ---
Gauss's law, that in terms of electric potential $\varphi$
can be expressed as Poisson equation.
In practice Poisson equation is used for correction
of ``potential'' part of electric field.

It is well known
that Particle-in-Cell (PIC) method in plasma simulations can be implemented
without solving Poisson equation for electric field correction.
Instead, we need the continuity equation (or charge conservation law) in finite
differences to be satisfyed.

There are a few methods for satisfying the continuity equation locally ---
for charge and current density associated with each particle,
Ref.~\cite{Morse-Nilson,Buneman,Vshivkov}.
For this purpose authors use special definition for the current density
wich is naturally connected with the change of charge density due to particle
motion.
Unfortunately, these methods are implemented only for
simple shapes of particles, for the zero- and the first-order form-factors.
We present the generalization of these methods,
valid for a big class of form-factors.
Also we have proved that the density decomposition is the only possible
linear procedure for defining the current density
associated with the motion of a particle.

There are another methods for incorporating Gauss's law into Maxwell solver
using usual definition of local current density,
see \cite{Langdon-92,Marder-87}.

Very detailed study of PIC method can be found
in~\cite{Langdon,Hockney,VA-book}.
The new construction will be usefull firstly for overdensed plasma simulation
with the paradigm of 'Clouds-in-Cell'~\cite{CIC}.

%%%%%%%%%%%%%%%%%%%%%%%%%%%%%%%%%%%%%%%%%%%%%%%%%%%%%%%%%%%%

\section{Continuity equation in finite differences}

Let us consider the local Maxwell solver, wich is equivalent to
Finite Difference Time Domain (FDTD) method~\cite{YEE}
\begin{eqnarray} \label{FDTD1}
\frac{{\bf E}^{n+1}-{\bf E}^n}{dt} =  \nabla^+ \times {\bf B}^{n+1/2} - {\cal J}^{n+1/2} \, ,
\\ \label{FDTD2}
\frac{{\bf B}^{n+1/2}-{\bf B}^{n-1/2}}{dt} = - \nabla^- \times {\bf E}^{n} \, ,
\\ \label{FDTD3}
\nabla^- \cdot{\bf E}^n = \rho^n \, ,
\\ \label{FDTD4}
\nabla^+ \cdot{\bf B}^{n+1/2} = 0 \, ,
\end{eqnarray}
combined with the particle mover
\begin{eqnarray} \label{MOVER1}
\frac{{\bf u}_\alpha^{n+1/2}-{\bf u}_\alpha^{n-1/2} }{dt} =
2\pi \frac{q_\alpha}{m_\alpha}
     \frac{m_e}{e}
\left( {\bf E}^n({\bf x}_\alpha^n,t) +
\frac{{\bf u}^n_\alpha}{\gamma_\alpha} \times{\bf B}^n({\bf x}_\alpha^n,t)\right) \, ,
\\ \label{MOVER2}
\frac{{\bf x}_\alpha^{n+1}-{\bf x}_\alpha^{n} }{dt} = \frac{{\bf u}_\alpha^{n+1/2}}{\gamma_\alpha^{n+1/2}} \, ,
\\ \gamma_\alpha=\left(1+({{\bf u}_\alpha})^2\right)^{1/2} \, .
\end{eqnarray}

Equations Eqs.(\ref{FDTD1}-\ref{FDTD4}) are discreetized Maxwell equations
and Eqs.(\ref{MOVER1}-\ref{MOVER2}) are leap-frog scheme for solving of
Newton-Lorentz equations. Here we use dimensionless variables defined
by transformations
$
t \rightarrow 2\pi \omega_0^{-1} t ,
{\bf x} \rightarrow \lambda_0{\bf x} ,
({\bf E},{\bf B}) \rightarrow (m_e c \omega_0/e) ({\bf E},{\bf B}) ,
$
where $m_e ,e$ --- electron mass and charge, $c$ --- speed of light,
$\omega_0$ and $\lambda_0$ --- some characteristic frequency and length
(e.g. the frequency and wavelength of incident EM radiation).
Index $n$ denotes integer time step and $\alpha$
stands for the number of a particle;
$dt,dx,dy,dz$ --- discreetization of time and space coordinates.

Different components of electromagnetic fields and charge density $\rho$
and current density ${\cal J}$ are defined on different grids,
\begin{eqnarray} \label{GRIDS}
{\bf E}=(E^1_{i,j+1/2,k+1/2},E^2_{i+1/2,j,k+1/2},E^3_{i+1/2,j+1/2,k}) ,
\quad
{\bf B}=(B^1_{i+1/2,j,k},B^2_{i,j+1/2,k},B^3_{i,j,k+1/2}) ,
\nonumber \\
\rho=\rho_{i+1/2,j+1/2,k+1/2} ,
\quad
{\cal J}=({\cal J}^1_{i,j+1/2,k+1/2},{\cal J}^2_{i+1/2,j,k+1/2},{\cal J}^3_{i+1/2,j+1/2,k}) ,
\end{eqnarray}
where $i,j,k$ are integers.
Discreet operators $\nabla^\pm$ in Eqs.(\ref{FDTD1}-\ref{FDTD4})
are vectors,
\begin{eqnarray}
\nabla^+ f_{i,j,k} =
\left(
\frac{f_{i+1,j,k}-f_{i,j,k}}{dx} ,
\frac{f_{i,j+1,k}-f_{i,j,k}}{dy} ,
\frac{f_{i,j,k+1}-f_{i,j,k}}{dz}
\right)
,
\nonumber \\
\nabla^- f_{i,j,k} =
\left(
\frac{f_{i,j,k}-f_{i-1,j,k}}{dx} ,
\frac{f_{i,j,k}-f_{i,j-1,k}}{dy} ,
\frac{f_{i,j,k}-f_{i,j,k-1}}{dz}
\right).
\end{eqnarray}
These operators have the next convenient properties
\begin{equation}
\nabla^- \times \nabla^+ = \nabla^+ \times \nabla^- = 0, \quad
\nabla^- \cdot \nabla^+ = \nabla^+ \cdot \nabla^- = \Delta ,
\end{equation}
where $\Delta$ is discreet Poisson operator in central differences,
\begin{equation}
\Delta f_{i,j,k}=
\frac{f_{i-1,j,k}-2 f_{i,j,k}+f_{i+1,j,k}}{dx^2}+
\frac{f_{i,j-1,k}-2 f_{i,j,k}+f_{i,j+1,k}}{dy^2}+
\frac{f_{i,j,k-1}-2 f_{i,j,k}+f_{i,j,k+1}}{dz^2} \, .
\end{equation}

Acting on the Eq.(\ref{FDTD1}) by $(\nabla^-\times)$
and on the Eq.(\ref{FDTD2}) by $(\nabla^+\times)$,
we obtain
\begin{eqnarray} \label{CONT-E}
\frac{\rho^{n+1}-\rho^n}{dt}+
\nabla^-\cdot{\cal J}^{n+1/2} = 0 \, ,
\\ \label{CONT-B}
\frac{\nabla^+{\bf B}^{n+1/2}-\nabla^+{\bf B}^{n-1/2}}{dt} = 0 \, .
\end{eqnarray}
It means that
if the continuity equation Eq.(\ref{CONT-E})
is fulfilled then the divergence of ${\bf E}$ is always equal to
charge density (Gauss's law), and if the initial discreet divergence
of ${\bf B}$ is zero then it remains zero forever.

Thus, for solving Maxwell equations
we need Eqs.(\ref{FDTD1}-\ref{FDTD2}) and Eq.(\ref{CONT-E})
with initial conditions
\begin{equation}
\nabla^- \cdot {\bf E} = \rho \quad \mbox{ and } \quad
\nabla^+ \cdot {\bf B}=0 \quad \mbox{ at } \quad t=0 \, .
\end{equation}

Let us consider
the continuity equation (or charge conservation law) in finite
differences
\begin{eqnarray}\label{CONTINUITY}
\frac{\rho_{i+1/2,j+1/2,k+1/2}^{n+1/2}-\rho_{i+1/2,j+1/2,k+1/2}^n}{dt}+
\frac{{\cal J}^1_{i,j+1/2,k+1/2}-{\cal J}^1_{i-1,j+1/2,k+1/2}}{dx}+
\nonumber \\
\frac{{\cal J}^2_{i+1/2,j,k+1/2}-{\cal J}^2_{i+1/2,j-1,k+1/2}}{dy}+
\frac{{\cal J}^3_{i+1/2,j+1/2,k}-{\cal J}^3_{i+1/2,j+1/2,k-1}}{dz}= 0 \, .
\end{eqnarray}
Further we will drop indices and modificators like $\pm 1/2$,
where it can not lead to an ambiguity.
The charge density $\rho$ is constructed from form-factors of separate
particles
\begin{equation}\label{RHO}
\rho_{i,j,k}=\sum_\alpha S_{i,j,k}(x_\alpha,y_\alpha,z_\alpha) ,
\end{equation}
where $S$ is the form-factor (or density) of a particle,
\begin{equation}\label{FORM-FACTOR}
S_{i,j,k}(x_\alpha,y_\alpha,z_\alpha) =
S(X_i-x_\alpha,Y_j-y_\alpha,Z_k-z_\alpha) ,
\end{equation}
$X_i,Y_j,Z_k$ denote coordinates of the grid,
$(x_\alpha,y_\alpha,z_\alpha)$ is the location of the particle
with number $\alpha$.
Here form-factor can be interpreted
as a charge density of a single particle. So the particle is considered
as it would be a charged cloud.
Form-factor must obey the rule of conservation of full charge
which leads to
\begin{equation}\label{WEIGHT}
\sum_{i,j,k} S_{i,j,k}(x_\alpha,y_\alpha,z_\alpha) = 1 ,
\end{equation}
where the sum is taken over all grid nodes.

%%%%%%%%%%%%%%%%%%%%%%%%%%%%%%%%%%%%%%%%%%%%%%%%%%%%%%%%%%%%

\section{Density decomposition}

Due to linearity of charge conservation law Eq.(\ref{CONTINUITY}),
it is sufficient to construct current density associated with
motion of a single particle.

Let us consider a single particle with form-factor Eq.(\ref{FORM-FACTOR})
and coordinates $(x,y,z)$.
We introduce vector $W$ as finite differences of the current density
associated with particle motion:
\begin{eqnarray}\label{W-DEF}
\nonumber
{\cal J}^1_{i,j,k}-{\cal J}^1_{i-1,j,k}=-\frac{dx}{dt} W^1_{i,j,k} \, ,
\\ \nonumber
{\cal J}^2_{i,j,k}-{\cal J}^2_{i,j-1,k}=-\frac{dy}{dt} W^2_{i,j,k} \, ,
\\
{\cal J}^3_{i,j,k}-{\cal J}^3_{i,j,k-1}=-\frac{dz}{dt} W^3_{i,j,k} \, .
\end{eqnarray}

Then according to charge conservation law,
we can write dropping grid indices,
\begin{equation}\label{CONSERVATION}
W^1+W^2+W^3=S(x+\Delta x,y+\Delta y,z+\Delta z)-S(x,y,z) .
\end{equation}
Here $(\Delta x,\Delta y,\Delta z)$ is 3-dimensional shift of the particle
due to motion.

Shift of the particle generates eight functions
\begin{eqnarray}\label{8-fun}
\nonumber
S(x,y,z),\quad
S(x+\Delta x,y,z), S(x,y+\Delta y,z), S(x,y,z+\Delta z),
\\ \nonumber
S(x+\Delta x,y+\Delta y,z), S(x+\Delta x,y,z+\Delta z), S(x,y+\Delta y,z+\Delta z),
\\
S(x+\Delta x,y+\Delta y,z+\Delta z) \, .
\end{eqnarray}
We will assume that vector $W$ and corresponding current density
linearly depends from these functions.
The base for this assumption is the following.
{\bf (1)}~We can consider the form-factor as charge density of the particle.
If form-factor amplitude is increasing, the current density
associated with a shift of the form-factor must increase proportionally.
{\bf (2)}~We can decompose any three-dimensional shift of form-factor $S(x,y,z)$
into three one-dimensional shifts:
\begin{eqnarray}
\nonumber
S(x+\Delta x,y+\Delta y,z+\Delta z)-S(x,y,z)=
\\ \nonumber
S(x+\Delta x,y,z)-S(x,y,z)
+
\\ \nonumber
S(x+\Delta x,y+\Delta y,z)-S(x+\Delta x,y,z)
+
\\
S(x+\Delta x,y+\Delta y,z+\Delta z)-S(x+\Delta x,y+\Delta y,z) .
\end{eqnarray}
Currents corresponding to each one-dimensional shift must be additive.

Let us formulate some conditions directly going form
the nature of vector $W$.

\begin{itemize}
\item[\bf 1.]
Vector $W^1_{i,j,k},W^2_{i,j,k},W^3_{i,j,k}$
is a decomposition of finite difference
$S_{i,j,k}(x+\Delta x,y+\Delta y,z+\Delta z)-S_{i,j,k}(x,y,z)$,
Eq.(\ref{CONSERVATION}).

\item[\bf 2.]
If some of shifts $\Delta x$, $\Delta y$,
$\Delta z$ iz zero, the corresponding component $W$
is also zero:

$\Delta x =0 \Rightarrow W^1=0$,
$\Delta y =0 \Rightarrow W^2=0$,
$\Delta z =0 \Rightarrow W^3=0$.

\item[\bf 3.]
If $S(x,y,z)$ is symmetrical with respect to permutation of $(x,y)$,
$S(x,y,z)=S(y,x,z)$
and $\Delta x=\Delta y$, then $W^1=W^2$.
The same property is assumed for symmetries with respect to permutations
of pairs $(x,z)$ and $(y,z)$.

\end{itemize}

\paragraph{Suggestion.}
{\it There is only one linear combination of eight functions Eq.(\ref{8-fun}),
each satisfying Eq.(\ref{WEIGHT}),
that is consistent with properties {\bf 1-3}:
}
\begin{eqnarray}\label{DECOMPOSITION}
%---------------------------------------------
\nonumber
W^1=
\frac{1}{3} S(x+\Delta x,y+\Delta y,z+\Delta z)
-
\frac{1}{3} S(x         ,y+\Delta y,z+\Delta z)
+
\\ \nonumber
+
\frac{1}{6} S(x+\Delta x,y         ,z+\Delta z)
-
\frac{1}{6} S(x         ,y         ,z+\Delta z)
+
\\ \nonumber
+
\frac{1}{6} S(x+\Delta x,y+\Delta y,z         )
-
\frac{1}{6} S(x         ,y+\Delta y,z         )
+
\\ \nonumber
+
\frac{1}{3} S(x+\Delta x,y         ,z         )
-
\frac{1}{3} S(x         ,y         ,z         )
\\ \nonumber
%---------------------------------------------
W^2=
\frac{1}{3} S(x+\Delta x,y+\Delta y,z+\Delta z)
-
\frac{1}{3} S(x+\Delta x,y         ,z+\Delta z)
+
\\ \nonumber
+
\frac{1}{6} S(x         ,y+\Delta y,z+\Delta z)
-
\frac{1}{6} S(x         ,y         ,z+\Delta z)
+
\\ \nonumber
+
\frac{1}{6} S(x+\Delta x,y+\Delta y,z         )
-
\frac{1}{6} S(x+\Delta x,y         ,z         )
+
\\ \nonumber
+
\frac{1}{3} S(x         ,y+\Delta y,z         )
-
\frac{1}{3} S(x         ,y         ,z         )
\\ \nonumber
%---------------------------------------------
W^3=
\frac{1}{3} S(x+\Delta x,y+\Delta y,z+\Delta z)
-
\frac{1}{3} S(x+\Delta x,y+\Delta y,z         )
+
\\ \nonumber
+
\frac{1}{6} S(x         ,y+\Delta y,z+\Delta z)
-
\frac{1}{6} S(x         ,y+\Delta y,z         )
+
\\ \nonumber
+
\frac{1}{6} S(x+\Delta x,y         ,z+\Delta z)
-
\frac{1}{6} S(x+\Delta x,y         ,z         )
+
\\
+
\frac{1}{3} S(x         ,y         ,z+\Delta z)
-
\frac{1}{3} S(x         ,y         ,z         )
\end{eqnarray}
\paragraph{Proof.} (Scenario). We can write all the properties {\bf 1-3}
in the form of linear equations with unknown coefficients
of eight functions.
Remembering Eq.(\ref{WEIGHT})
we can obtain additional equations on coefficients
taking sum over all grid points $(i,j,k)$ from each linear
combination for $W$.
Solving 10 linear equations for all $S$, we will find all the coefficients.
Of course, not all eight values Eq.(\ref{8-fun}) are independent.
We have six independend variables $x,y,z,\Delta x,\Delta y,\Delta z$,
so in the most general case only six values $S$
can be also independend, for example, excluding $S(x,y,z)$ and
$S(x+\Delta x,y+\Delta y,z+\Delta z)$.
Among all possible solutions we must left only one, which
doesn't assume special numerical values for excluded functions. $\Box$

Taking into account boundary conditions for the current of one particle
(vanishing of the current density at nodes far from the form-factor domain),
and using Eq.({\ref{WEIGHT}}) we obtain:
\begin{eqnarray}\label{BND}
\nonumber
\sum_{i} W^1_{i,j,k}=0 \, ,
\\ \nonumber
\sum_{j} W^2_{i,j,k}=0 \, ,
\\
\sum_{k} W^3_{i,j,k}=0 \, .
\end{eqnarray}

Two systems Eq.(\ref{DECOMPOSITION}) and Eq.(\ref{BND})
define the density decomposition. Solving Eq.(\ref{W-DEF})
with natural boundary condition we obtain the current density
associated with a single particle motion.

The condition Eq.(\ref{BND}) can be easily satisfyed
if form-factor have a property of inheritance in decreasing of the dimension,
i.e. if sum of form-factor over any dimension is again form-factor
but of lower dimension.
Formally, it means
\begin{equation}
S^{(2D)}_{i,j}(x,y)=\sum_{k} S^{(3D)}_{i,j,k}(x,y,z) ,
\end{equation}
where $S^{(2D)}_{i,j}$ doesn't depend on $z$ and obeys Eq.(\ref{WEIGHT})
automatically.

There is a big and widely used in PIC codes
class of form-factors that have a property of inheritance:
all form-factors that are the products of one-dimensional form-factors,
\begin{equation}\label{CLASS}
S^{3D}_{i,j,k}(x,y,z) = S^{1D}_i(x) S^{1D}_j(y) S^{1D}_k(z) .
\end{equation}
Here we use the same symbol for (probably) different one-dimensional form-factors,
each of them must satisfy conservation of full charge, Eq.(\ref{WEIGHT}).

It can be easily proved that density decomposition Eq.(\ref{DECOMPOSITION})
along with Eq.(\ref{CLASS}) is the generalization of techniques
proposed in~\cite{Morse-Nilson,Buneman,Vshivkov}.

%%%%%%%%%%%%%%%%%%%%%%%%%%%%%%%%%%%%%%%%%%%%%%%%%%%%%%%%%%%%

\section{Computing of the current with second-order polynomial form-factor}

In this section we present an algorithm for density decomposition
in the case of second-order piecewise-polynomial form-factor
and discuss a problem of dimension reduction.

Let us consider well-known one-dimensional form-factor
\begin{eqnarray}
\nonumber
S^{(1D)}_{i}(x) = \frac{3}{4} - (X_i-x)^2 \, &,&
\\
S^{(1D)}_{i\pm 1}(x) = \frac{1}{2} \left( \frac{1}{2} \mp (X_i-x) \right)^2 \, &,&
\quad |X_i-x|<1/2 \, ,
\end{eqnarray}
which is the second-order spline. The particle is bell-shaped.
The correspondent 3-dimensional form-factor is Eq.(\ref{CLASS}).

Now we can formulate a scenario for computing the current density
based on density decomposition Eq.(\ref{DECOMPOSITION}).
Suppose we consider a code that uses
Finite Difference Time Domain (FDTD) technique~\cite{YEE},
where electromagnetic fields and current density are
defined on different regular grids.
Here we do not pretend to show optimized or fastest algorithm.
\begin{itemize}
\item[\bf 1.]
Prepare 15-component array ${\sf S0}$ containing
one-dimensional form-factors
corresponding to particle coordinates $({\sf x0,y0,z0})$ with respect to
the grid of the charge density $\rho$:
\begin{eqnarray}
\nonumber
{\sf S0}(i,1)=S^{(1D)}_i({\sf x0}) \, , i=-2,2 \, ,
\\ \nonumber
{\sf S0}(j,2)=S^{(1D)}_j({\sf y0}) \, , j=-2,2 \, ,
\\
{\sf S0}(k,3)=S^{(1D)}_k({\sf z0}) \, , k=-2,2 \, , 
\end{eqnarray}
Really, components ${\sf S0}(-2,m)$ and ${\sf S0}(2,m)$ are zero,
but we need these additional components for further calculations.

The actual 3-dimensional form-factor is 27-component array
\begin{equation}
S^{(3D)}(i,j,k)={\sf S0}(i,1)*{\sf S0}(j,2)*{\sf S0}(k,3) \, .
\end{equation}

\item[\bf 2.]
Using ${\sf S0}$ or precomputed $S^{(3D)}$, compute the force acting
on the particle. Here we can use fields spatially
averaged to the grid of $\rho$ or compute additional form-factors for each
type of grid.
Advance particle and compute new particle coordinates
$({\sf x1,y1,z1})$. Note here that particle shift in any direction
must be smaller or equal than grid step in this direction,
\begin{equation}\label{PART-LIMIT}
{\sf x1-x0} \le dx ,\quad {\sf y1-y0} \le dy ,\quad {\sf z1-z0} \le dz .
\end{equation}

\item[\bf 3.]
Using new particle coordinates compute a new array ${\sf S1}$
containing new form-factors:
\begin{eqnarray}
\nonumber
{\sf S1}(i,1)=S^{(1D)}_i({\sf x1}) \, , i=-2,2 \, ,
\\ \nonumber
{\sf S1}(j,2)=S^{(1D)}_j({\sf y1}) \, , j=-2,2 \, ,
\\
{\sf S1}(k,3)=S^{(1D)}_k({\sf z1}) \, , k=-2,2 \, .
\end{eqnarray}
Components ${\sf S1}(-2,m)$ and ${\sf S1}(-2,m)$ are not
zero in general, because of particle motion.
If conditions Eq.(\ref{PART-LIMIT}) are satisfyed,
the array ${\sf S1}(i,m)$ doesn't have non-zero components out of
$i =-2,2$.

\item[\bf 4.]
Compute auxiliary array of differences of new and old form-factors:
\begin{eqnarray}
\nonumber
{\sf DS}(i,1)={\sf S1}(i,1)-{\sf S0}(i,1) \, , i=-2,2 \, ,
\\ \nonumber
{\sf DS}(j,2)={\sf S1}(j,2)-{\sf S0}(j,2) \, , j=-2,2 \, ,
\\
{\sf DS}(k,3)={\sf S1}(k,3)-{\sf S0}(k,3) \, , k=-2,2 \, .
\end{eqnarray}
It is possible to use ${\sf S1}$ for storage of differences.

\item[\bf 5.]
Compute 125*3-component array containing density decomposition
${\sf W}(i,j,k,m)$, in accordance with Eq.(\ref{DECOMPOSITION}).
We need so many componets
because we have current whose components are defined on different regular
grids (in FDTD technique).
\begin{eqnarray} \label{A-5}
\lefteqn{}&&
\nonumber
{\sf W}(i,j,k,1)={\sf DS}(i,1)*\bigl(
             {\sf S0}(j,2)*{\sf S0}(k,3)
+\frac{1}{2}*{\sf DS}(j,2)*{\sf S0}(k,3)+ 
\\ \nonumber
&&
+\frac{1}{2}*{\sf S0}(j,2)*{\sf DS}(k,3)
+\frac{1}{3}*{\sf DS}(j,2)*{\sf DS}(k,3) \bigr)	\, ,
\\ \nonumber
&&
{\sf W}(i,j,k,2)={\sf DS}(j,2)*\bigl(
             {\sf S0}(i,1)*{\sf S0}(k,3)
+\frac{1}{2}*{\sf DS}(i,1)*{\sf S0}(k,3)+
\\ \nonumber
&&
+\frac{1}{2}*{\sf S0}(i,1)*{\sf DS}(k,3)
+\frac{1}{3}*{\sf DS}(i,1)*{\sf DS}(k,3) \bigr)	\, ,
\\ \nonumber
&&
{\sf W}(i,j,k,3)={\sf DS}(k,3)*\bigl(
             {\sf S0}(i,1)*{\sf S0}(j,2)
+\frac{1}{2}*{\sf DS}(i,1)*{\sf S0}(j,2)+
\\
&&
+\frac{1}{2}*{\sf S0}(i,1)*{\sf DS}(j,2)
+\frac{1}{3}*{\sf DS}(i,1)*{\sf DS}(j,2) \bigr)	\, .
\end{eqnarray}
Of course, this computation is easy to optimize.

\item[\bf 6.]
Compute three components of the current density ${\cal J}^1,{\cal J}^2,{\cal J}^3$
associated with motion of the particle,
using Eq.(\ref{W-DEF}) and boundary condition (there is no current in nodes
far from particle location),
\begin{eqnarray} \label{A-6}
\nonumber
{\cal J}^1_{i,j,k}-{\cal J}^1_{i-1,j,k}=-{\sf Q} \frac{dx}{dt} {\sf W}(i,j,k,1) \, ,
\\ \nonumber
{\cal J}^2_{i,j,k}-{\cal J}^2_{i,j-1,k}=-{\sf Q} \frac{dy}{dt} {\sf W}(i,j,k,2) \, ,
\\
{\cal J}^3_{i,j,k}-{\cal J}^3_{i,j,k-1}=-{\sf Q} \frac{dz}{dt} {\sf W}(i,j,k,3) \, ,
\end{eqnarray}
where ${\sf Q}$ is the charge of the particle.

\item[\bf 7.]
Add computed contribution from the single particle to
array of the current density.

\end{itemize}
As this algorithm uses only simple polynomes,
its accuracy is equivalent to the accuracy of the last digit
of numerical representation (e.g. $10^{-8}$ in SINGLE PRECISION
4-BYTE data or $10^{-17}$ in DOUBLE PRECISION 8-BYTE data).

Suppose we have two-dimensional problem,
when all the variables depend on $(x,y)$ only.
In this case density decomposition Eq.(\ref{DECOMPOSITION}) provides
only two first components of the current density.
How to construct the third one, in consistency with the rest?
The simplest idea is to derive the third component from 3-dimensional
case by reducing the dimension.
We can imagine chaines of infinite number of particles along $z$-axise.
Being projected into $(x,y)$-plane these $N$ chaines produces
$N$ 2-dimensional particles.
Then we can do averaging over z-axise.
As a result we will obtain first two components of the current density
in accordance with Eq.(\ref{DECOMPOSITION}),
and the third component.

In  the particular case
of the above algorithm we must change formulae of items {\bf 5}
and {\bf 6} in the following way:
%\begin{itemize}
%\item[\bf 5$^{(2D)}$.]

\begin{eqnarray}
\lefteqn{}&&
\nonumber
{\sf W}(i,j,1)={\sf DS}(i,1)*\bigl(
             {\sf S0}(j,2)
+\frac{1}{2}*{\sf DS}(j,2) \bigr)	\, ,
\\ \nonumber
&&
{\sf W}(i,j,2)={\sf DS}(j,2)*\bigl(
             {\sf S0}(i,1)
+\frac{1}{2}*{\sf DS}(i,1) \bigr)	\, ,
\\ \nonumber
&&
{\sf W}(i,j,3)=
             {\sf S0}(i,1)*{\sf S0}(j,2)
+\frac{1}{2}*{\sf DS}(i,1)*{\sf S0}(j,2)+
\\
&&
+\frac{1}{2}*{\sf S0}(i,1)*{\sf DS}(j,2)
+\frac{1}{3}*{\sf DS}(i,1)*{\sf DS}(j,2)	\, .
\end{eqnarray}

%\item[\bf 6$^{(2D)}$.]

\begin{eqnarray}
\nonumber
{\cal J}^1_{i+1,j}-{\cal J}^1_{i,j}=-{\sf Q} \frac{dx}{dt} {\sf W}(i,j,1) \, ,
\\ \nonumber
{\cal J}^2_{i,j+1}-{\cal J}^2_{i,j}=-{\sf Q} \frac{dy}{dt} {\sf W}(i,j,2) \, ,
\\
{\cal J}^3_{i,j}=-{\sf Q} {\sf V_z} {\sf W}(i,j,3) \, ,
\end{eqnarray}
where ${\sf V_z}$ is the third component of particle velocity.

%\end{itemize}
As one can see these formulae have an obvious connection
with 3D-case, Eqs.(\ref{A-5}-\ref{A-6}).

%%%%%%%%%%%%%%%%%%%%%%%%%%%%%%%%%%%%%%%%%%%%%%%%%%%%%%%%%%%%

\section{Conclusion}

In this paper we have developed a construction for
a current density, which exactly satisfy the charge conservation law
and is valid for a wide class of form-factors.
It is shown that this construction is the only
allowed by very natural conditions derived from the properties of
the current density.
An algorithm in the case of second-order polynomial form-factor
is presented.
One can see that this method is not restricted by special Maxwell solver,
but uses only discreetized continuity equation.
These teqnique was implemented by author and D.V.Sokolov
in three-dimensional and two-dimensional PIC codes.

The author is glad to thank Dmitry Sokolov for collaboration,
Prof. Vitaly A. Vshivkov and Dr. Hartmut Ruhl
for useful discussion.

The author is pleased to thank Prof. Francesco Pegoraro
and Prof. Giuseppe Bertin for support.

This work was prepared in Scuola Normale Superiore in Pisa
and supported by
Istituto Nazionale per la Fisica della Materia, Italy
and by
Russian Fond for Basic Research (No.98-02-16298).

\end{document}